\documentclass[cits,a4paper]{PoS}
\newcommand{\half}{\textstyle{1\over2}}

\PoS{PoS(LAT2005)255}
\title{Large-N QCD at strong transverse lattice gauge coupling}
\ShortTitle{Large-N QCD at strong transverse lattice gauge coupling}
\author{\speaker{Apoorva D. Patel}\footnote{%
        Submitted to PoS on 24 November 2005.}\\
        Centre for High Energy Physics, Indian Institute of Science,
        Bangalore-560012, India\\
        E-mail: \email{adpatel@cts.iisc.ernet.in}}
\author{Raghunath Ratabole\\
        The Institute of Mathematical Sciences,
        C.I.T. Campus, Taramani, Chennai-600113, India\\
        E-mail: \email{raghu@imsc.res.in}}
\abstract{
We had previously obtained an integral equation for mesons in transverse
lattice QCD, in the limit of large number of colours and strong transverse
lattice gauge coupling~\cite{RAGHU}.
This equation is a generalisation of the 't~Hooft equation~\cite{THOOFT}, 
by inclusion of the spin degrees of freedom. We analyse this equation
to extract spectral properties and light-front wavefunctions of mesons.
We also extend the method to study baryon properties in the same limit.}
\FullConference{XXIIIrd International Symposium on Lattice Field Theory\\
                25-30 July 2005\\
                Trinity College, Dublin, Ireland}

\begin{document}

\section{The formalism}

One of us had pointed out that transverse lattice QCD (i.e. QCD with
two continuous dimensions on light-front and two transverse dimensions
on lattice), can be solved analytically in a closed form in the combined
$N\rightarrow\infty$ and strong transverse gauge coupling limits \cite{PATEL}.
Although this limiting theory differs from real QCD, we hope that the
results will be particularly useful for understanding deep-inelastic
scattering structure functions dominated by valence quarks. In this case,
the scaling behaviour implies that the transverse momenta do not contribute
to the leading order, and so distorting them using a lattice may not be a
severe drawback as long as we treat the light-front components exactly.

We study the limit of QCD defined by the action
($g$ is held fixed as $N\rightarrow\infty$):
\begin{eqnarray}
S &=& a_\perp^2 \sum_{x_\perp}\int d^2x~ \Big[ -{N \over 4g^2} \sum_{\mu\nu a}
      F_{\mu\nu}^a(x) F^{\mu\nu a}(x)
  ~+~ \overline{\psi}(x) \Big( i\sum_\mu \gamma^\mu \partial_\mu
               - \sum_\mu \gamma^\mu A_\mu - m \Big) \psi(x) \nonumber\\
  &+& {\kappa \over 2a_\perp}\sum_n \Big\{
      \overline{\psi}(x) (1+i\gamma^n) U_n(x) \psi(x+\hat{n}a_\perp)
   +  \overline{\psi}(x+\hat{n}a_\perp) (1-i\gamma^n) U_n^\dag(x)
                                          \psi(x) \Big\} \Big] ~.
\end{eqnarray}
Here $\mu,\nu$ label the light-front directions, and $n$ labels the lattice
directions. In the $g_\perp\rightarrow\infty$ limit, the transverse lattice
spacing $a_\perp = O(\Lambda_{QCD})$. For the fermions, we follow the
Wilson prescription in the transverse directions. The anisotropy parameter
$\kappa$ has to be determined non-perturbatively by demanding as much
restoration of rotational symmetry as possible, and the metric is
\begin{equation}
\half\{\gamma_\alpha,\gamma_\beta\} = g_{\alpha\beta} = \left(\matrix{
0 & 1 &  0 &  0 \cr
1 & 0 &  0 &  0 \cr
0 & 0 & -1 &  0 \cr
0 & 0 &  0 & -1 \cr}\right) ~.
\end{equation}
In analysing this theory, the order of limits is important, because various
limits do not commute. To obtain the correct phase of the theory, we first
let $g_\perp\rightarrow\infty$, then $N\rightarrow\infty$, and then
$m\rightarrow0$.

Starting with the above action, we eliminate the gauge degrees of freedom
completely by exact functional integrations over $A^-$ (after choosing
$A^+=0$) and $U_n(x)$. We then trade off the fermion fields in favour of
non-local boson fields, and obtain an effective action in terms of
$\sigma_{\alpha\beta}(x,y)\equiv\overline{\psi}_\alpha(x)\psi_\beta(y)$
\cite{RAGHU}. The stationary point value of this effective action,
$V_{\rm eff}(\overline{\sigma};J)$, yields the generating functional for
the connected Green's functions, as $N\rightarrow\infty$.

For Wilson fermions. the projection operator structure of the fermion hopping
term simplifies the formulae, and the results are simple modifications of
those for the 't~Hooft model. The transverse tadpole insertions renormalising
the quark mass vanish, and the chiral limit of the theory remains at $m=0$
\cite{RAGHU}. The chiral condensate, obtained using split-point regularisation
and operator formulation \cite{BURKARDT}, is
\begin{equation}
\langle\overline{\psi}\psi\rangle_{4d}
~=~ {2 \over a_\perp^2}\langle\overline{\psi}\psi\rangle_{2d}
~\mathop{\longrightarrow}\limits_{m\rightarrow0}~
-{N \over a_\perp^3}\sqrt{g^2 \over 3\pi} ~.
\end{equation}

\section{Meson states}

The wavefunction for the meson state with spin-parity structure $\Gamma$
is defined as
\begin{equation}
\phi_\Gamma(p,q) = \langle \overline{\psi}(p-q) \Gamma \psi(q)
                 | {\rm Meson}_\Gamma (p) \rangle ~.
\end{equation}
It satisfies a homogeneous Bethe-Salpeter equation with two types of
quark-antiquark interactions: gluon exchange in the longitudinal direction,
and bilinear fermion hopping in the transverse directions. In the reference
frame with momentum components $p=(p^+=1,p^-=M^2/2,p_\perp=0)$, the
interactions are independent of the ``$-$'' and ``$\perp$'' components,
and the Bethe-Salpeter equation is easily projected on to the light-front:
\begin{equation}
\Phi_\Gamma(q^+ \equiv x)
= \int dq^-~\frac{d^2q_\perp}{(2\pi)^2}~\phi_\Gamma(p,q) ~.
\end{equation}
For a quark-antiquark pair of masses $m_1$ and $m_2$, we obtain
($\beta \equiv g^2/\pi a_\perp^2$, P$\equiv$principal value) \cite{RAGHU},
\begin{eqnarray}
\label{integraleqn}
\mu^2(x)\Phi(x) &\equiv& \left[M^2-\frac{m_1^2-\beta}{x}-\frac{m_2^2-\beta}{1-x}\right]\Phi(x) \nonumber\\
&=& \frac{1}{2x(1-x)}\left[\frac{m_1^2}{2x}\gamma^+ +x\gamma^- +m_1\right] \\
&\times& \int_0^1 \frac{dy}{2\pi}\Big\{ -\frac{g^2}{a_\perp^2}\mathrm{P}\Big[\frac{1}{(x-y)^2}\Big]\gamma^+\Phi(y)\gamma^+
~+~ \frac{\kappa^2}{a_\perp^2}\Big[2\Phi(y) + \sum_{n}\gamma^n\Phi(y)\gamma^n\Big] \Big\} \nonumber\\
&\times& \left[\frac{\mu^2(x)}{2}\gamma^+ +\frac{m_2^2}{2(1-x)}\gamma^+ +(1-x)\gamma^- -m_2\right] .\nonumber
\end{eqnarray}

\subsection{Symmetry consequences}

Eq.(\ref{integraleqn}) is a $16$-component matrix integral equation in Dirac
space, and physical meson states have to be obtained by diagonalising it.
Discrete symmetries of the action allow block-diagonalisation of
Eq.(\ref{integraleqn}) to four blocks of $4$ components each.
Writing meson wavefunctions in a basis that is the direct product
of Clifford algebra bases in continuum and lattice directions,
\begin{equation}
\Phi = \sum_{C,L}\Phi_{C;L}\Gamma^{C;L} ~,~~~~
\Gamma^{C;L} \in \left\{1,\gamma^+,\gamma^-,\half[\gamma^+,\gamma^-]\right\}
\otimes
\left\{1,\gamma^{n_1},\gamma^{n_2},\half[\gamma^{n_1},\gamma^{n_2}]\right\} ~,
\end{equation}
we have $\sum_n\gamma^n\Gamma^{C;L}\gamma^n \propto \Gamma^{C;L}$
for each value of $L$. Thus the lattice index ``$L$'' can be used
as the block label, and the exact degeneracy of $L=n_1$ and $L=n_2$
blocks implies that only three of the four blocks are independent.

Without the transverse lattice dynamics, Eq.(\ref{integraleqn}) reduces to
the 't~Hooft equation \cite{THOOFT}, and the limit $\kappa\rightarrow0$ is
smooth. The non-singular transverse contribution proportional to $\kappa^2$
represents a colour singlet quark-antiquark pair hopping from one light-front
to the next. (Leaving out the $\gamma$-matrices, the form of this contribution
is similar to the fermion-antifermion annihilation diagram for the massive
Schwinger model \cite{BERGKNOFF}.) The strong transverse gauge coupling
limit produces a tight binding $\delta$-function constraint during the
hopping, so that the transverse interaction is a wavefunction at the origin
effect and the meson orbital angular momentum $L_z$ vanishes. With spin-half
quarks, the meson helicities are restricted to $0,\pm1$, and the allowed
spin-parity quantum numbers for the mesons are $J^P=0^\pm,1^\pm$.

Under the exchange of the quark and the antiquark, the meson wavefunction
transforms as $E_{q\overline{q}}\Phi(x;m_1,m_2)=C\Phi^T(1-x;m_2,m_1)C^{-1}$,
where $C$ is the charge conjugation operator. Eq.(\ref{integraleqn}) is
invariant under this exchange operation. Although parity is an exact
symmetry of our formalism, it is not manifest because the light-front is
not invariant under parity transformation. The $(x \leftrightarrow 1-x)$
part of $E_{q\overline{q}}$ can be associated with parity, however, and
this parity symmetry is exact in every block labeled by $L$. As a result,
the eight possible spin-parity quantum numbers are distributed in to the
four blocks as two states of opposite parity in each block. In conventional
notation, $\{\pi,a_1(0)\}\in\Phi_{C;n_1n_2}$, the degenerate pair
$\{\rho(n),a_1(n)\}\in\Phi_{C;n}$, and $\{\rho(0),\sigma\}\in\Phi_{C;1}$.

Restricted to the finite box $x\in[0,1]$, the spectrum of $M^2$ is purely
discrete as in the 't~Hooft model \cite{THOOFT}, and the meson states can
be labeled by a radial excitation quantum number $n=1,2,3\ldots$ in each
block. Since the lowest state meson wavefunction $\Phi^{(n=1)}_{-;L}$ is
symmetric, the exchange symmetry alternates with $n$, and the quark and
the antiquark have opposite intrinsic parities, the meson states have
parity $P=(-1)^n$ in each block.

\subsection{Behaviour in the chiral limit}

The singular part of the interaction kernel in Eq.(\ref{integraleqn})
is the same as in the 't-Hooft model, and depends only on the components
$\Phi_{-;L}$ for each value of $L$. The behaviour of the solutions in
certain limiting situations, therefore, can be obtained using the same
methods as for the 't~Hooft model \cite{THOOFT,BERGKNOFF,CCG,EINHORN,COLEMAN}.

The singular part can be separated using the projection,
$\gamma^+\Phi\gamma^+ = \sum_L\Phi_{-;L} (2\Gamma^L\otimes\gamma^+)$,
\begin{equation}
M^2\Phi_{-;L}(x) = \left[\frac{m_1^2}{x}+\frac{m_2^2}{1-x}\right]
   \Phi_{-;L}(x) + \beta\int_0^1 dy~\mathrm{P}
   \left[\frac{\Phi_{-;L}(x)-\Phi_{-;L}(y)}{(x-y)^2}\right] + \chi_{-;L}(x) ~,
\end{equation}
\begin{equation}
\chi_{-;L}(x) = \frac{\kappa^2}{2\pi a_\perp^2}\int_0^1 dy \cdot\cases{
    4\Phi_{+;1}(y) -{2m_1m_2 \over x(1-x)}\Phi_{-;1}(y)   ~,\cr 
    -2\Phi_{+;n}(y) -{m_1m_2 \over x(1-x)}\Phi_{-;n}(y)
    -({m_1 \over x}+{m_2 \over 1-x})\Phi_{1;n}(y)
    +({m_1 \over x}-{m_2 \over 1-x})\Phi_{+-;n}(y)        ~,\cr 
    2({m_1 \over x}-{m_2 \over 1-x})\Phi_{1;n_1n_2}(y)
    -2({m_1 \over x}+{m_2 \over 1-x})\Phi_{+-;n_1n_2}(y) ~.\cr 
}
\end{equation}
For finite norm solutions, the integrals appearing in $\chi_{-;L}$ have to
be finite. The solutions $\Phi_{-;L}$ therefore vanish at the boundaries
as $x^{\beta _{1}}$ and $(1-x)^{\beta _{2}}$, and the exponents can be
determined using an ansatz $\Phi_{-;L} \sim x^{\beta_1}(1-x)^{\beta_2}$.

The equations simplify in the chiral limit. $\chi_{-;n_1 n_2}$
vanishes when $m=0$, and consequently $\Phi^{(n)}_{-;n_1 n_2}(m=0)$
coincide with the corresponding solutions of the 't Hooft equation.
In particular, the lightest meson is massless and has the wavefunction
$\Phi^{(n=1)}_{-;n_1 n_2}(m=0)=1$. This is the pseudoscalar Goldstone
boson of the theory, with the decay constant
\begin{equation}
(f_\pi)_{4d} ~=~ \sqrt{2} (f_\pi)_{2d}
~\mathop{\longrightarrow}\limits_{m\rightarrow0}~
{1 \over a_\perp}\sqrt{2N \over \pi} ~.
\end{equation}
Asymptotically for large $n$, the masses and the wavefunctions behave as
\begin{equation}
n\gg 1 :~~ \Phi_{-;n_1 n_2}^{(n)}(x)\simeq \sqrt{2}\sin (n\pi x) ~,~~
           M_n^2 \simeq n\pi g^2/a_\perp^2 ~.
\end{equation}
With the physical values $N=3$ and $f_\pi\simeq 130$ MeV,
we estimate the cut-off as $\pi/a_\perp\simeq 300$ MeV.
Fitting the slope of the $\pi-\pi(1300)-\pi(1800)$ trajectory to its
asymptotic behaviour, then gives the gauge coupling $g^2/4\pi\simeq 2.3$.
With these parameters, the chiral condensate turns out to be
$\langle\overline{\psi}\psi\rangle\simeq -(165~{\rm MeV})^3$.

The masses of mesons in the other spin-parity blocks are shifted due to
the non-singular part of the interaction kernel. The solutions have to
be determined numerically, and the parameter $\kappa$ can then be fixed
by making the helicity$=0,\pm$ states for $J=1$ mesons as degenerate as
possible.

\section{Baryon states}

Baryons are semi-classical solitons in the $N\rightarrow\infty$ limit.
In this scenario, each valence quark can be considered to be moving in
the common Hartree potential provided by the other $N-1$ valence quarks
(sea quarks drop out in the $N\rightarrow\infty$ limit) \cite{HARTREE}.
This potential is static and of finite range, and carries colour opposite
to that of the valence quark. The potential experienced by a valence quark
bound to a heavy antiquark has the same features, although it may have a
different spatial dependence, and several techniques used to study the
heavy quark effective theory can be applied to the baryon case as well.

The baryon wavefunctions are completely antisymmetric in colour, and so
fully symmetric in space, spin and flavour indices. Furthermore, in the
Hartree approximation, the ground state baryons have all the valence
quarks in the same lowest state of the potential. The total wavefunction
is thus the product of identical single-particle wavefunctions; it is
fully symmetric in space and satisfies $I=J$. The baryons are solutions
of the bosonised effective action, with
\begin{equation}
Q(x^+) \equiv a_\perp^2\sum_{x_\perp}\int dx^-
\overline{\psi}(x,x_\perp) \gamma^+\psi(x,x_\perp) = {\rm const}.
\end{equation}
So in the single baryon sector, the $N\rightarrow\infty$ stationary point
of the effective action has to satisfy $a_\perp^2\sum_{x_\perp}\int dx^-
{\rm tr} (\overline{\sigma}_{\alpha\beta}(x,x) \gamma^+_{\alpha\beta}) = N$.
Extremisation of $V_{\rm eff}(\sigma;J=0)$ leads to
\begin{eqnarray}
1 &=& i\overline{\sigma}^T(x,y) (i\partial\!\!\!\!\!/-m) \delta^{(2)}(x-y)
   - {ig^2\over 2} |x^- - y^-| ~\overline{\sigma}^T(x,y)\gamma^+\overline{\sigma}^T(y,x)\gamma^+ \nonumber\\
  &-& i\delta^{(2)}(x-y) \sum_n \big[ \overline{\sigma}^T(x,x) G(\overline{R}) (1-i\gamma_n) \overline{\sigma}^T(x-n,x-n) (1+i\gamma_n) \\
  & & \qquad\qquad\quad          +~ \overline{\sigma}^T(x,x) (1+i\gamma_n) \overline{\sigma}^T(x+n,x+n) G(\overline{R}) (1-i\gamma_n)
                               \big] ~, \nonumber
\end{eqnarray}
where we have chosen units to set $a_\perp=1$ for simplicity, and
\begin{equation}
G(R) = {\kappa^2 \over 2(1+\sqrt{1+\kappa^2 R})} ~,~~~~
R = - (1-i\gamma_n) \sigma^T(x,x) (1+i\gamma_n) \sigma^T(x+n,x+n) ~.
\end{equation}

The meson stationary point is translationally invariant,
\begin{equation}
\overline{\sigma}^T_{B=0}(x,y) = -i \int {d^2p \over (2\pi)^2}
  ~ {1 \over p\!\!\!\!/-m-\Sigma_{B=0}(p)+i\epsilon}
  ~ e^{ip\cdot(x-y)} \delta_{x_\perp y_\perp} ~,~~~~
\Sigma_{B=0} (p) = - {g^2\gamma^+ \over 2\pi p^+} ~.
\end{equation}
With $\overline{\sigma}_{B=0}(x,x) \propto 1$ and $\overline{R}_{B=0}=0$,
the transverse lattice dynamics doesn't contribute to it. On the contrary,
the baryon stationary point is not translationally invariant, and we
decompose \cite{RAJEEV}
\begin{equation}
\label{valquark}
\overline{\sigma}_{B=1}(x,y) = \overline{\sigma}_{B=0}(x,y)
+ \delta_{x_\perp y_\perp} f^\ast(x^-) \gamma^+ f(y^-) ~,~~~~
\int dx^- |f(x^-)|^2 = {\textstyle{1 \over 4}} ~,
\end{equation}
in the Hartree approximation. The stationary point equation then yields
(in momentum space),
\begin{eqnarray}
0 &=& \int {d^2k\over(2\pi)^2} |\tilde{f}(k)|^2 \gamma^+\gamma^- (k\!\!\!\!/ - \widetilde{m} + {g^2\gamma^+ \over \pi k^+}) \\
  &+& 4g^2\gamma^+ \int {d^2p\over(2\pi)^2} {d^2q\over(2\pi)^2} {d^2k\over(2\pi)^2}
      \mathrm{P}\Big[{1\over(p^+)^2}\Big] \tilde{f}^\ast(q) \tilde{f}(k) \tilde{f}^\ast(k+p) \tilde{f}(q+p) ~. \nonumber
\end{eqnarray}
Here the transverse lattice dynamics contributes only through the
wavefunction at the origin effect, and renormalises the quark mass
as $\widetilde{m}=m + 8G(0)\langle\overline{\psi}\psi\rangle/N$.
Extracting the $\gamma^+$-component,
\begin{equation}
\int {dk^+ \over 4\pi^2k^+} |\tilde{f}(k^+)|^2
+ \int {dp^+ \over 2\pi} {dq^+ \over 2\pi} {dk^+ \over 2\pi}
  \mathrm{P}\Big[{1\over(p^+)^2}\Big] \tilde{f}^\ast(q^+) \tilde{f}(k^+) \tilde{f}^\ast(k^++p^+) \tilde{f}(q^++p^+) = 0 ~.
\end{equation}
To obtain the valence quark density in the baryon, $4|\tilde{f}(k^+)|^2$
with $k^+\ge0$, this nonlinear integral equation has to be solved
numerically, as in the case of the 't Hooft model \cite{RAJEEV}.

Alternatively, the baryon number constraint can be incorporated in the
functional integral using a constant temporal Abelian background field,
i.e. the chemical potential $\mu$ (see e.g. \cite{BARYON}),
\begin{equation}
\delta(Q - NB) = \int [D\mu] ~\exp\big[i(Q-NB)\mu] ~.
\end{equation}
Functional integration at a fixed chemical potential shifts the self-energy
of the quark propagator,
\begin{equation}
S(p) = {i \over p\!\!\!\!/ - m - \Sigma_{\mu\ne0} (p) + i\epsilon} \delta_{x_\perp y_\perp} ~,~~~~
\Sigma_{\mu\ne0} (p) = - \left({g^2 \over 2\pi p^+} + \mu\right) \gamma^+ ~.
\end{equation}
Indeed, this structure justifies the decomposition in Eq.(\ref{valquark}).

\end{document}